# Anomalous critical point behavior in dilute magnetic semiconductor $(Ca,Na)(Zn,Mn)_2Sb_2$

Shuang Yu[1,2], Xinyu Liu[3], Guoqiang Zhao[1,2], Yi Peng[1], Xiancheng Wang[1,2], Jianfa Zhao[1,2], Wenmin Li[1,2], Zheng Deng[1,2,a)], Jacek K. Furdyna[3], Y. J. Uemura[4] and Changqing Jin[1,2,5,b)]

[1] *Beijing National Laboratory for Condensed Matter Physics, and Institute of Physics, Chinese Academy of Sciences, Beijing 100190, China*
[2] *School of Physics, University of Chinese Academy of Sciences, Beijing 100190, China*
[3] *Department of Physics, University of Notre Dame, Notre Dame, Indiana 46556, USA*
[4] *Department of Physics, Columbia University, New York, New York, 10027, USA.*
[5] *Materials Research Lab at Songshan Lake, 523808 Dongguan, China*

[a)]Electronic mail: dengzheng@iphy.ac.cn

[b)]Electronic mail: Jin@iphy.ac.cn

## Abstract

In this paper we report successful synthesis and magnetic properties of $(Ca,Na)(Zn,Mn)_2Sb_2$ as a new ferromagnetic dilute magnetic semiconductor (DMS). In this DMS material the concentration of magnetic moments can be controlled independently from the concentration of electric charge carriers that are required for mediating magnetic interactions between these moments. This feature allows us to separately investigate the effect of carriers and of spins on the ferromagnetic properties of this new DMS alloy, and particularly of its critical ferromagnetic behavior. We use modified Arrott plot analysis to establish critical exponents $\beta$, $\gamma$, and $\delta$ for this alloy. We find that at low Mn concentrations (< 10 at.%), it is governed by short-range 3D-Ising-like behavior, with experimental values of $\beta$, $\gamma$, and $\delta$ very close to theoretical 3D-Ising values of 0.325, 1.24, and 4.815. However, as the Mn concentration increases, this DMS material exhibits a mixed-phase behavior, with $\gamma$ retaining its 3D-Ising characteristics, but $\beta$ crossing over to longer-range mean-field behavior.

## 1. INTRODUCTION

Ferromagnetic (FM) ordering in dilute magnetic semiconductors (DMSs) is based on carrier-mediated exchange interaction between spins localized on magnetic ions [1-3]. Controlling the state of the carriers then offers the ability to tune magnetic properties of such DMSs, thus leading to potential applications in spintronic devices [4-7]. However, the detailed mechanisms of FM interactions in DMSs are still in question, even in the case of the (Ga,Mn)As system that has been extensively studied over nearly two decades as a model DMS material. Although the mean-field-like Zener model is widely accepted for describing this and related DMS materials, many studies reveal inconsistent results based on their critical behavior in the proximity of the Curie temperature ($T_C$) [8-11]. Specifically, short-range as well as long-range exchange interactions (e.g., Heisenberg or Ising models, and mean-field models) have been proposed for describing magnetic properties in these systems [8,11-13].

These conflicting descriptions could result from consequences of the complicated non-equilibrium growth of (Ga,Mn)As [5]. For example, magnetic properties of this alloy depend strongly on the specific location of Mn ions in the (Ga,Mn)As crystal lattice, since Mn ions at interstitial sites are ferromagnetically inactive [14]. The aforementioned difficulty in (Ga,Mn)As is mainly due to the limited capability of $Mn^{2+}$ to substitute for $Ga^{3+}$ [5]. More important, one should note that in (Ga,Mn)As the $Mn^{2+}$ ions are simultaneously the source of magnetic spins and of charge carriers, making it difficult to distinguish between the roles of spins and carriers on the resulting ferromagnetism of the material.

Recently a new class of DMS materials has been discovered – e.g.,

$Li_{1+x}(Zn,Mn)As$, termed “111-type”, and $(Ba,K)(Zn,Mn)_2As_2$, termed “122” – in which carrier doping and spin doping are independent from each other, thus overcoming the difficulty just described [15-21]. For example, in $(Ba,K)(Zn,Mn)_2As_2$ the isovalent substitution of $Mn^{2+}$ for $Zn^{2+}$ enables one to achieve high-Mn solubility (up to ~15% or more) in thermodynamic equilibrium, while substitution of $K^+$ for $Ba^{2+}$ produces charge carriers (holes) [22,23]. One can then *independently* control the concentrations of localized spins and charge carriers, and study their respective roles in determining magnetic properties of such DMSs – an important advantage over (Ga,Mn)As and related materials, in which such separation of the two key elements of ferromagnetism is not possible [24]. In particular, these features make the new DMSs – with their independent carrier and spin doping – especially attractive for testing and distinguishing between the various mechanisms proposed for magnetic interactions in DMS systems [25,26].

In this work we use modified Arrott plots (MAPs) to study the critical behavior, in the proximity of the Curie temperature, of the DMS alloy $(Ca,Na)(Zn,Mn)_2Sb_2$, a system with spatially and electronically separated sources of charges and spins. In this alloy the substitution of $Na^+$ for $Ca^{2+}$ produces holes, and isovalent substitution of $Mn^{2+}$ for $Zn^{2+}$ introduces localized spins. In the present paper we will focus on the finding that, by fixing the Na doping level at 10% and increasing Mn concentrations from 5% to 15%, we observe an anomalous critical behavior in $(Ca,Na)(Zn,Mn)_2Sb_2$ with highest Mn concentration, where a mixed magnetic state occurs in which exchange interaction reflects both the 3D-Ising and mean-field features.

## 2. EXPERIMENT

Polycrystalline specimens of $(Ca,Na)(Zn,Mn)_2Sb_2$ were synthesized by solid state reaction of high purity elements. Stoichiometric ratios of starting materials were mixed thoroughly and pressed into pellets. All processes were conducted in high-purity argon atmosphere to avoid oxidation of air-sensitive starting materials. The pellets were sealed in tantalum tubes with 1 bar of Argon. The Ta tubes were enclosed in evacuated quartz tubes, and the samples were heated to 700 ℃ for 12 hours. The products were then reground, pelleted, and sintered at 750 ℃ for an additional 12 hours.

The crystal structure of the resulting specimens was characterized by powder x-ray diffraction (PXRD) using a Philips X'pert diffractometer at room temperature. DC magnetic susceptibility was measured between 2 and 300 K using a Superconducting Quantum Interference Device (SQUID) magnetometer. A Physical Properties Measurement System (PPMS) was used for electrical transport measurements. To ensure that all specimens used in Arrott plot were initially magnetized, their isothermal magnetizations were measured after the samples were warmed up well above the Curie temperature $T_C$.

## 3. RESULTS

### 3.1 Crystal structure

The alloy $(Ca,Na)(Zn,Mn)_2Sb_2$ discussed in this paper crystallizes in hexagonal $CaAl_2Si_2$-type structure (space group *P-3m1*) that consists of alternating $(Zn,Mn)_2Sb_2$ and (Ca,Na) layers, as shown in Figure 1(a) [27]. The PXRD patterns for $(Ca_{1-x}Na_x)(Zn_{0.85}Mn_{0.15})_2Sb_2$ ($x$ = 0, 0.05, 0.1, and 0.15) are shown in Figure S1. As plotted in Figure 1(b), the unit cell volume increases linearly with increasing Mn doping level, but decreases with increasing Na doping, suggesting successful chemical doping by both Na and Mn.

### 3.2 Basic magnetic and transport properties

Although the parent phase $CaZn_2Sb_2$ itself has a relatively high carrier density ($n_p$ ~ $10^{19}cm^{-3}$) [28], samples doped only by Mn are clearly paramagnetic (Figure S2). On the other hand, samples co-doped with Na and Mn show a transition from the paramagnetic (PM) to the FM state at low temperatures, as seen in Figures 2(a) and 2(b). The Curie temperature in these specimens is controlled by both Na and Mn doping levels, indicating intrinsic ferromagnetism rather than FM clusters. Interestingly, $(Ca_{0.9}Na_{0.1})(Zn_{0.90}Mn_{0.10})_2Sb_2$, the sample with the highest Curie temperature in our series ($T_C$ = 10.8 K), is obtained with intermediate Na and Mn concentrations. Further increases in Na or Mn doping are observed to reduce $T_C$.

The susceptibility above $T_C$ is well fitted with the Curie-Weiss law, $\chi \sim C/(T-\theta)$, where $C$ is the Curie constant, and $\theta$ is the Weiss temperature. A positive value of $\theta$ indicates FM interactions between Mn spins. As an example, for $(Ca_{0.9}Na_{0.1})(Zn_{0.90}Mn_{0.10})_2Sb_2$ we observe $\theta$ = 12.8 K, consistent with ferromagnetic/paramagnetic transition temperature ($T_C$), indicated by the minimum in $dM(T)/dT$ *vs* $T$ curve shown in Figure 2(c).

Systematic measurements of the resistivity $\rho(T)$ show the following effects of Na- and Mn-doping. As seen in Figures 2(d) and 2(e), the resistivity decreases with increasing Na concentration both in samples doped only by Na and in those co-doped by Na and Mn, clearly indicating an increase in carrier concentration due to $Na_{ca}$ substitution. In contrast, resistivity increases with increasing Mn concentration in samples doped only by Mn and in samples co-doped by Na and Mn (Figures 2(f) and 2(g)), which we ascribe to an increase of magnetic scattering centers. Additionally, the latter two figures show a local maximum in $\rho(T)$, indicating a paramagnetic-to-ferromagnetic (PM/FM) transition. This is due to the fact that in the PM range carriers are scattered by random spin fluctuation, which is reduced as the sample orders ferromagnetically at $T_C$. This behavior is consistent with other DMS materials that exhibit FM order.

In the low-temperature range the resistivity shows a maximum near the Curie temperature, along with a large negative magnetoresistance typically observed in FM materials (see Figure S3). Figure 2(h) shows Hall resistivity ($\rho_{xy}(H)$) below and above $T_C$ for $(Ca_{0.9}Na_{0.1})(Zn_{0.9}Mn_{0.1})_2Sb_2$. At $T$ = 2 K $\rho_{xy}(H)$ clearly manifests anomalous Hall effect (AHE), again providing strong evidence for intrinsic ferromagnetism of the specimen. $\rho_{xy}(H)$ loop of 2K shows a coercive field of about 300 Oe which is close to corresponding $M(H)$ loop. Carrier concentration calculated from the linear portion of the $\rho_{xy}(H)$ curve measured at 2 K in the high-field range is $n_p \sim 1.4\times10^{20}$ cm$^{-3}$ ($p$-type). At 50 K, well above $T_C$, $\rho_{xy}$ is proportional to the applied field, and we obtain $n_p \sim 1.5\times10^{20}$ cm$^{-3}$, in good agreement with the 2 K result.

Compared with $n_p \sim 3.1\times10^{19}$cm$^{-3}$ in undoped $CaZn_2Sb_2$, the 10% Na doping has thus resulted in a prominent increase of hole concentration in the material. One should note here that the value of $n_p$ observed in $(Ca_{0.9}Na_{0.1})(Zn_{0.9}Mn_{0.1})_2Sb_2$ is comparable to that typically seen in other DMSs in which ferromagnetism is mediated by mobile carriers, such as (Ga,Mn)As, Li(Zn,Mn)As, $(Ba,K)(Zn,Mn)_2As_2$, as well as the analogue of the material presently investigated, $(Ca,Na)(Zn,Mn)_2As_2$ [15,16,29,30,31,32].

### 3.3 Critical behavior of $(Ca,Na)(Zn,Mn)_2Sb_2$ with low Mn concentration

We now analyze the critical behaviors of two samples with low Mn concentrations, $(Ca_{0.9}Na_{0.1})(Zn_{0.95}Mn_{0.05})_2Sb_2$ and $(Ca_{0.9}Na_{0.1})(Zn_{0.9}Mn_{0.1})_2Sb_2$, using the Arrott plot method. Generally, critical exponents and critical temperatures can be easily determined from the Arrott–Noakes equation of state [33,34]:

$$(H/M)^{1/\gamma} = a(T\text{-}T_C)/T_C + bM^{1/\beta}. \qquad (1)$$

Here $a$ and $b$ are constants, while $\beta$ is the critical exponent relevant to spontaneous magnetization and $\gamma$ is the critical exponent associated with initial magnetic susceptibility above transition temperature $T_C$. In Figure 3 we construct modified Arrott plots for $(Ca_{0.9}Na_{0.1})(Zn_{0.90}Mn_{0.10})_2Sb_2$ using critical exponents $\beta$ and $\gamma$ theoretically predicted for mean-field ($\beta$ = 0.5, $\gamma$ = 1), 3D-Heisenberg ($\beta$ = 0.365, $\gamma$ = 1.386), 3D-Ising ($\beta$ = 0.325, $\gamma$ = 1.24), and tricritical mean-field ($\beta$ = 0.25, $\gamma$ = 1) models.

For a diluted magnetic semiconductor, disorder is expected to truncate the critical region significantly [35]. Thus a narrow critical region of $\varepsilon \equiv \frac{|T-T_C|}{T_C} < 0.3$ is used to analyze the critical behaviors of all samples. In Figure 3(a) we present the isothermal magnetization $M(H)$ of $(Ca_{0.9}Na_{0.1})(Zn_{0.90}Mn_{0.10})_2Sb_2$ measured from 0 to 4 T near the transition from FM to PM state, measured below and above at Curie temperature $T_C$. Figure 3(b) shows an Arrott plot obtained by theoretical mean field theory with critical exponents values of $\beta = 0.5$, $\gamma = 1$ for the same series of temperatures near $T_C$. Clearly the curves do not form a series of parallel lines in the high field region, indicating that the standard mean-field model is not appropriate for $(Ca_{0.9}Na_{0.1})(Zn_{0.90}Mn_{0.10})_2Sb_2$. We then construct modified Arrott plots for the same set of data using critical exponents for the 3D-Heisenberg model ($\beta = 0.365$, $\gamma = 1.386$), 3D-Ising model ($\beta = 0.325$, $\gamma = 1.24$) and tricritical mean-field model ($\beta = 0.25$, $\gamma = 1$) for this sample (Figures 3(c)-3(e)). These three models yield nearly-straight and parallel lines in the high field region. In order to determine which one of them is the best for the determination of critical exponents, we calculated the so called normalized slope ($NS$) defined at the critical point as $NS(T) = S(T)/S(T_C)$. The $NS$ versus temperature for all four models is shown in Figure 3(f). Note that, if the modified Arrott plots show a series of absolute parallel lines, $NS(T)$ of the most satisfactory model should have the constant value of 1.0 irrespective of temperature. As shown in Figure 3(f), one can see that the 3D-Ising model is most suitable for $(Ca_{0.9}Na_{0.1})(Zn_{0.90}Mn_{0.10})_2Sb_2$. Thus the 3D-Ising model parameters are used as initial values to obtain accurate critical exponents by iteration. Modified Arrott plots for another low-Mn sample, $(Ca_{0.9}Na_{0.1})(Zn_{0.95}Mn_{0.05})_2Sb_2$, are shown in Figure S4, where we also find that the 3D-Ising model is most suitable, confirming our conclusion that samples of this allow with low Mn doping are best described by this model.

As is well known, in the vicinity of the FM-PM transition spontaneous magnetization $M_s$ and initial susceptibility $\chi_0$ can be described by the following equations [33], obtained from the divergence of correlation length according to universal scaling laws:

$$M_s(T) = M_0(-\varepsilon)^{\beta}, \qquad \varepsilon<0,\ T<T_C, \qquad (2)$$

$$\chi_0^{-1}(T) = (h_0/M_0)\varepsilon^{\gamma}, \qquad \varepsilon>0,\ T>T_C, \qquad (3)$$

$$M = DH^{1/\delta}, \qquad \varepsilon = 0,\ T = T_C. \qquad (4)$$

Here $\varepsilon \equiv (T\text{-}T_C)/T_C$; $\beta$, $\gamma$, $\delta$ are critical exponents; and $h_0/M_0$ and $D$ are critical amplitudes. By extrapolating the linear part in the high field limit in Figure 3(d) (3D Ising model) to the *y*- and *x*-axes, we obtain spontaneous magnetization $M_s$ and initial inverse susceptibility $\chi_0^{-1}$ at various temperatures. The $M_s(T)$ (below $T_C$) and $\chi_0^{-1}(T)$ (above $T_C$) curves are fitted with Eqs. (2) and (3), yielding the first set of critical exponents ($\beta'$, $\gamma'$ and $T_C'$). Then a modified Arrott plot $M^{1/\beta'}$ *vs* $(H/M)^{1/\gamma'}$ is reconstructed by using the extracted values. With the reconstructed modified Arrott plot, a second set of critical exponents ($\beta''$, $\gamma''$ and $T_C''$) can be obtained. After several iterations, $\beta$, $\gamma$ and $T_C$ converge to the values around 0.352±0.023, 1.344±0.013 and 8.8±0.1 K (Figures 4(a) and 4(b)), respectively. Lower panel in Figure 4(b) shows that a log-log plot of $M(H)$ curve at 8.8K is a straight line in the high field region. According to Eq. (4), the slope of the fitted line gives $\delta$ = 4.821±0.001, in reasonable agreement with the value $\delta$ = 4.818 obtained from the Widom scaling relation, $\delta = 1 + \gamma/\beta$ [33,36]. This self-consistency in the value of $\delta$ suggests that critical exponents obtained in this analysis are reliable. Following similar analysis, critical exponents of $\beta$ = 0.322 ± 0.010, $\gamma$ = 1.385 ± 0.011 and $\delta$ = 5.078±0.021, and $T_C$ = 6.0±0.1 K are obtained for our second low-Mn sample, $(Ca_{0.9}Na_{0.1})(Zn_{0.95}Mn_{0.05})_2Sb_2$ (see Figure S5).

### 3.4 Critical behavior of $(Ca,Na)(Zn,Mn)_2Sb_2$ with high Mn concentration

We now turn to the results obtained for sample with higher Mn content, $(Ca_{0.9}Na_{0.1})(Zn_{0.85}Mn_{0.15})_2Sb_2$. Figure 5(a) shows plots of isothermal magnetization of this sample as a function of magnetic field measured around its $T_C$. Figures 5(b)-5(e) show the magnetization results obtained for this specimen, plotted with theoretical critical exponents (see Table I) for the standard mean-field, 3D-Heisenberg, 3D-Ising, and tricritical mean-field models, respectively [37]. It is surprising that only the standard mean-field model in Figure 5(b) displays a reasonable Arrott plot behavior, in the form of a fan of concave and convex curves around the transition temperature $T_C$. On the other hand, the plots obtained with 3D-Heisenberg, 3D-Ising, and tricritical mean-field models cannot lead to reasonable values of spontaneous magnetization $M_s$ and initial inverse susceptibility $\chi_0^{-1}$ around $T_C$, since all these curves are concave in only one direction. In Figure 5(f), we plot *NS* as a function of temperature for all four models. All values of *NS*(*T*) are strongly temperature-dependent, indicating that none of these models are appropriate for determining critical exponents. Nevertheless, the standard mean-field model parameters are used to initiate the iteration process for obtaining critical exponents for this specimen, since only plots for that model seen in Figure 5(b) appear reasonable.

To obtain precise critical exponents, iteration method is again used for analysis, as has been done in the preceding section, except that for this specimen we start from the standard mean-field results (Figure 5(b)) [38]. After several iterations the critical exponents $\beta$ and $\gamma$, as well as $T_C$, are found to converge to values of 0.483±0.012, 1.134±0.043 and 7.5±0.1 K, respectively, as shown in Figure 6(a). The final set of modified Arrott plots shows a series of nearly perfectly parallel lines at high field, as shown in Figure 6(c). Figure 6(b) shows the $M(H)$ curve on a log-log scale at 7.5 K. By fitting its slope, we obtain $\delta$ = 3.064±0.006, in reasonable agreement with the calculated value of $\delta$ = 3.348 using the Widom scaling relation. The obtained values of critical exponents and $T_C$ are thus self-consistent, and appear to accurately characterize the critical behavior in our high-Mn-concentration specimen. However, these critical exponents do not agree with critical exponents predicted for any single model, indicating that the estimated critical exponents for $(Ca_{0.9}Na_{0.1})(Zn_{0.85}Mn_{0.15})_2Sb_2$ specimen do not belong to any conventional universality class, falling between values predicted by the 3D-Ising model and the mean-field model. From this we infer that at this Mn concentration magnetic properties of $(Ca,Na)(Zn,Mn)_2Sb_2$ are best described as a mixed magnetic phase with mean field and 3D-Ising behaviors.

## 4. DISCUSSION

The values of the critical exponents $\beta$, $\gamma$ and $\delta$, as well as transition temperature $T_C$, obtained from modified Arrott Plots, are presented in the Table I along with the theoretically predicted values for different models. We observed continuous changes of critical exponents with increasing Mn concentrations for $(Ca_{0.9}Na_{0.1})(Zn_{1-y}Mn_y)_2Sb_2$ specimens with $y$ = 0.05, 0.1 and 0.15. From the Table I, we see that the values of the exponents for these two samples with lower Mn concentration are similar to those of 3D-Ising model, suggesting that the exchange interaction in $(Ca_{0.9}Na_{0.1})(Zn_{1-y}Mn_y)_2Sb_2$ may exhibit strong uniaxial anisotropy associated with its layered structure. On the other hand, the critical exponents for our sample with higher Mn concentration, $(Ca_{0.9}Na_{0.1})(Zn_{0.85}Mn_{0.15})_2Sb_2$, do not belong to any of the conventional universality classes. For this Mn concentration, we obtained critical exponent $\beta$ of 0.483±0.012, which is close to the theoretical value for the standard mean-field model, while exponent $\gamma$ obtained for this sample is 1.134±0.043, which is close to the value for the 3D-Ising model (see Table I).

The observed difference in $\beta$ from the value predicted by the 3D-Ising model may result from two possibilities: (i) extended exchange interaction, i.e., interactions beyond nearest neighbors; and (ii) magnetic inhomogeneities (i.e., disorder) in the system, which mixes phases with short-range and long-range exchanges. Additional studies are therefore still needed to gain further understanding of the behaviors revealed by the observed results. Helpful in this may be the trends which our result indicate, i.e., that the value of $\beta$ systematically increases and the value of $\gamma$ decreases with increasing Mn concentration.

Returning to our data, it is noteworthy that highest Curie temperature ($T_C$ = 8.8 K) occurs in the sample with an intermediate Mn content, $(Ca_{0.9}Na_{0.1})(Zn_{0.90}Mn_{0.10})_2Sb_2$, but drops to 7.5 K in the $(Ca_{0.9}Na_{0.1})(Zn_{0.85}Mn_{0.15})_2Sb_2$ specimen. This is probably caused by increasing antiferromagnetic superexchange between near-neighbor Mn cations as Mn doping attains sufficiently high levels. This speculation is also consistent with the $M(T)$ curves observed in experiments, indicating that the effective magnetic moment per Mn ion decreases as Mn concentration increases, as shown in Figure 2(a). In addition, we note that the resistivity of the samples also increases with Mn concentration, revealing an inhomogeneous nature of $(Ca_{0.9}Na_{0.1})(Zn_{1-y}Mn_y)_2Sb_2$ system, such that the magnetic inhomogeneity or disorder reduce the mean-free path of the holes that mediate magnetic interaction between magnetic moments in the system. Interestingly, very similar behavior has been observed in the case of II-VI based DMSs alloys (e.g., $Zn_{1-x}Mn_xSe$), where one observes a clear reduction of the effective Mn moment with increasing Mn concentration, such that beyond a certain point the value of magnetic moment per Mn ion actually *decreases* with further increase of Mn doping, in effect exhibiting a crossover-like behavior between samples with low-Mn and high-Mn content [39].

## 5. SUMMARY AND CONCLUSIONS

In summary, we have successfully synthesized a new ferromagnetic DMS material, $(Ca_{1-x}Na_x)(Zn_{1-y}Mn_y)_2Sb_2$, where spin and charge doping can be independently controlled. The ferromagnetic properties of this alloy, including its Curie temperature, are then controlled by the Na and Mn doping levels. This provides the opportunity of investigating the roles of the two dopants separately, a major advantage over the well-known III-V based DMS systems in which these roles cannot be separated. Detailed analysis of magnetization data obtained on $(Ca_{1-x}Na_x)(Zn_{1-y}Mn_y)_2Sb_2$ using the modified Arrott plots near the Curie temperature provided important insights on critical exponents $\beta$, $\gamma$, and $\delta$ for this alloy. Our analysis indicates that at low Mn concentrations (for values of $y \leq 0.10$) this DMS system displays clear 3D-Ising characteristics. For higher values of $y$, however, although the critical exponent $\gamma$ continues to retain its 3D-Ising characteristics, the value of $\beta$ significantly increases, crossing over to longer-range mean-field behavior, thus resulting in a mixed-phase 3D-Ising/3D-Heisenberg ferromagnet. We suggest that this cross-over of magnetic behavior, including the conspicuous increase in the value of $\beta$, may result from the increased role of intrinsic inhomogeneities or of increased antiferromagnetic coupling between Mn pairs, since both effects act to weaken short-range interactions between spins.

Interestingly, we observe that the increase in Mn content beyond $y \approx 0.10$ is accompanied by a decrease of the Curie temperature. Such behavior indicates an onset of antiferromagnetic interactions between Mn moments, similar to the behavior observed in II-VI based DMSs [39]. One should note that studies and conclusions of the type described in this paper are made particularly convenient in $(Ca_{1-x}Na_x)(Zn_{1-y}Mn_y)_2Sb_2$ because of the ability to independently control the concentrations of charge carrier (by varying the Na content) and of magnetic moments (by varying the concentration of Mn), which enables the investigator to distinguish the roles which the two dopants play in determining the magnetic properties of this ferromagnetic system.

**Acknowledgments**

This work was financially supported by National Key R&D Program of China (No. 2017YFB0405703), Ministry of Science and Technology of China

(2018YFA03057001) & National Natural Science Foundation of China through the research projects (11534016 and 11974407). Z.D. also acknowledges support of the Youth Innovation Promotion Association of CAS (No. 2020007).

## Figure and Table captions

**FIG. 1.** (a) Schematic diagram of (Ca,Na)(Zn,Mn)$_2$Sb$_2$ structure; (b) Unit cell volume change as a function of Mn and Na doping levels. Note that the Zn/Mn layers are the sites of magnetic spins, and that carriers originate in the Ca/Na layers.

**FIG. 2.** (a) $M(T)$ for $(Ca_{1-x}Na_x)(Zn_{1-y}Mn_y)_2Sb_2$ ($x$, $y$ = 0.05, 0.1 0.15) at $H$ = 500 Oe, and (b) $M(H)$ curves measured at 2 K. Insert in (a): a typical example of Curie-Weiss fitting for the $(Ca_{0.9}Na_{0.1})(Zn_{0.9}Mn_{0.1})_2Sb_2$ data. (c) $dM(\mathrm{T})/dT$ for $(Ca_{0.9}Na_{0.1})(Zn_{1-y}Mn_y)_2Sb_2$, with $y$ = 0.05, 0.1 and 0.15. $T_C$ are marked in the figure. Temperature dependence of resistivity $\rho(T)$ at zero field for (d) $(Ca_{1-x}Na_x)Zn_2Sb_2$ ($x$ = 0, 0.05, 0.1), (e) $(Ca_{1-x}Na_x)(Zn_{0.9}Mn_{0.1})_2Sb_2$ ($x$ = 0.05, 0.1, 0.15), (f) $Ca(Zn_{1-y}Mn_y)_2Sb_2$ ($y$ = 0, 0.05, 0.1, 0.15) and (g) $(Ca_{0.9}Na_{0.1})(Zn_{1-y}Mn_y)_2Sb_2$ ($y$ = 0.05, 0.1, 0.15). (g) Hall measurement results for $(Ca_{0.9}Na_{0.1})(Zn_{0.9}Mn_{0.1})_2Sb_2$ below and above the Curie temperature.

**FIG. 3.** (a) $M(H)$ curves for $(Ca_{0.9}Na_{0.1})(Zn_{0.9}Mn_{0.1})_2Sb_2$ at various temperatures ranging from 5.8-12.1 K with a step 0.2 or 0.3 K. The plots of $H/M^{1/\gamma}$ versus $M^{1/\beta}$ for the $(Ca_{0.9}Na_{0.1})(Zn_{0.9}Mn_{0.1})_2Sb_2$ specimen for (b) Standard mean-field model, (c) 3D-Heisenberg model, (d) 3D-Ising model, and (e) Tricritical mean-field model. (f) Temperature dependence of the normalized slopes, obtained from Figs. 3(b)-3(e).

**FIG. 4.** (a) Temperature variations of $M_s$ and $\chi_0^{-1}$ along with the fits (solid lines) following Eqs. (2) and (3), which give the values of the critical exponents and $T_C$ as mentioned in the plot. (b) Fitting results of $M(H)$ at 8.8 K obtained with Eq. (4). (c) Reconstructed modified Arrott-plot using obtained critical exponents, as discussed in the text.

**FIG. 5.** (a) $M(H)$ curves for $(Ca_{0.9}Na_{0.1})(Zn_{0.85}Mn_{0.15})_2Sb_2$ at various temperatures ranging from 5.5 to 9.8 K, in steps of 0.3 K. (b) – (e): Plots of $H/M^{1/\gamma}$ versus $M^{1/\beta}$ for this specimen for (b) standard mean-field model, (c) 3D-Heisenberg model, (d) 3D-Ising model, and (e) tricritical mean-field model, respectively. (f) Temperature dependence of the normalized slopes, obtained from Figs. 5(b) - 5(e).

**FIG. 6.** (a) Temperature variation of $M_s$ and $\chi_0^{-1}$, along with fits (solid lines) following Eqs. (2) and (3), which give the values of critical exponents and $T_C$ shown in the plot. (b) Fitting results of $M(H)$ at 7.5 K obtained with Eq. (4). (c) Reconstructed modified Arrott-plot using critical exponents shown in (a), as discussed in the text.

**Table. I.** Comparison of critical exponents of $(Ca_{0.9}Na_{0.1})(Zn_{1-y}Mn_y)_2Sb_2$ with earlier reports on different theoretical models. MAP stands for modified Arrott plots.

**FIG. 1.**

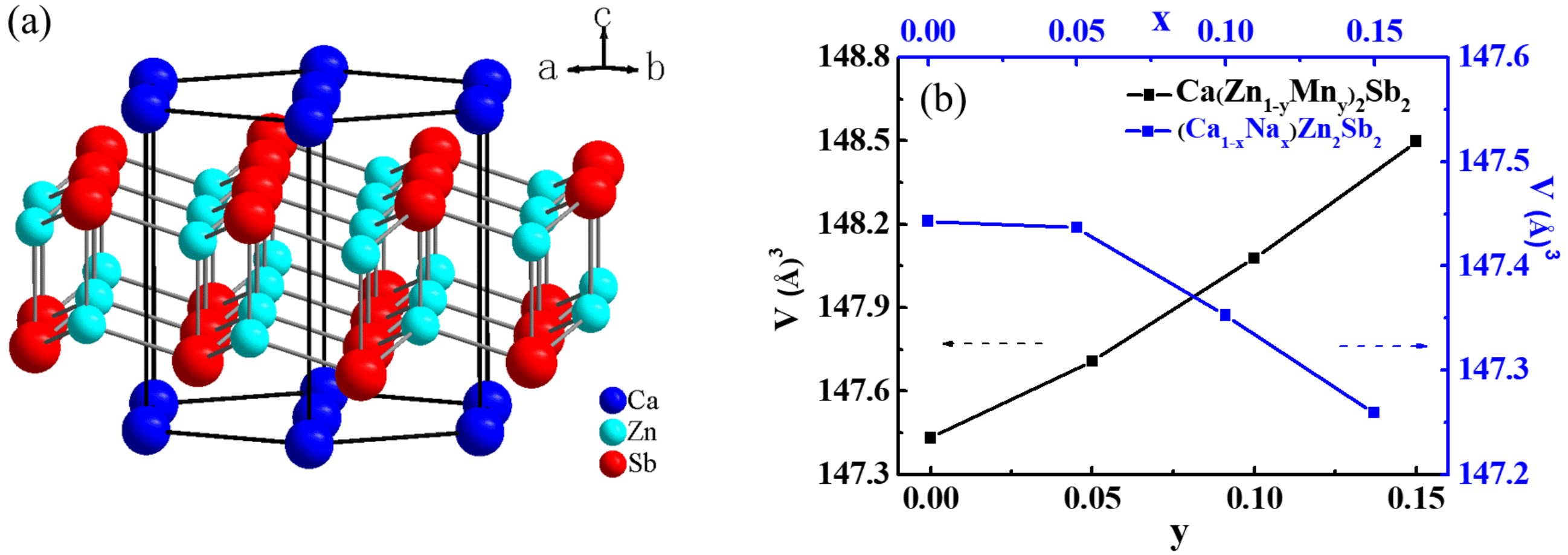

FIG. 2.

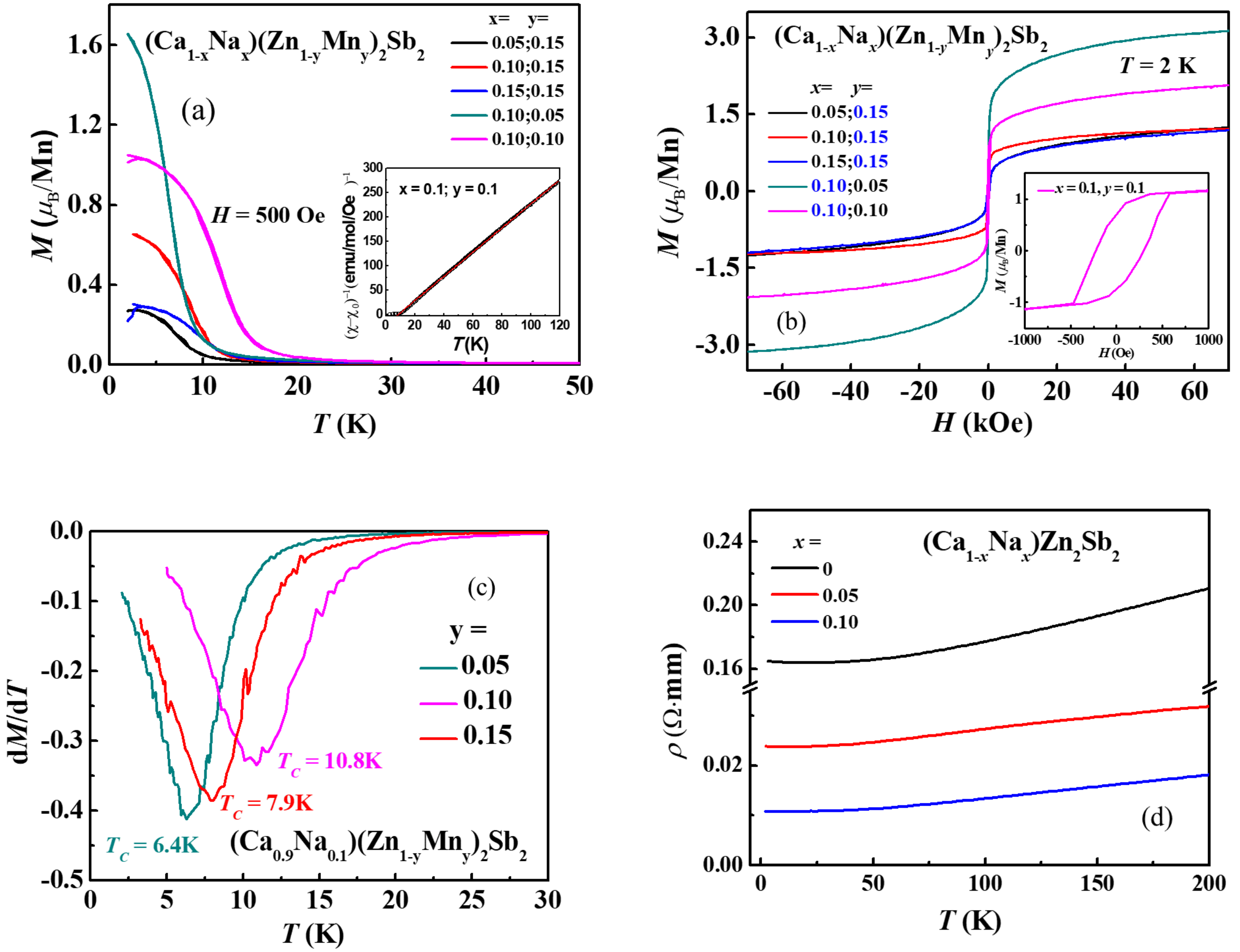

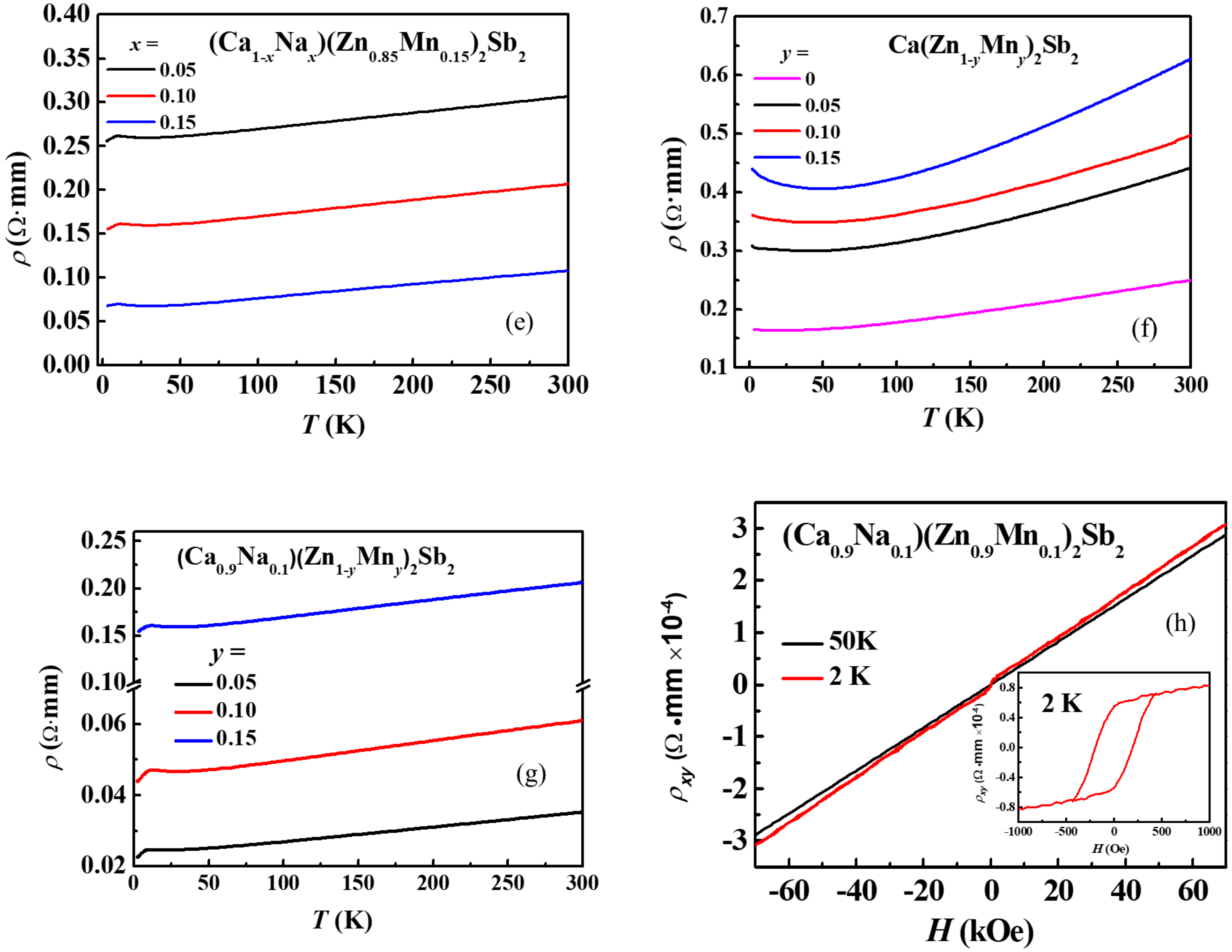

FIG. 3.

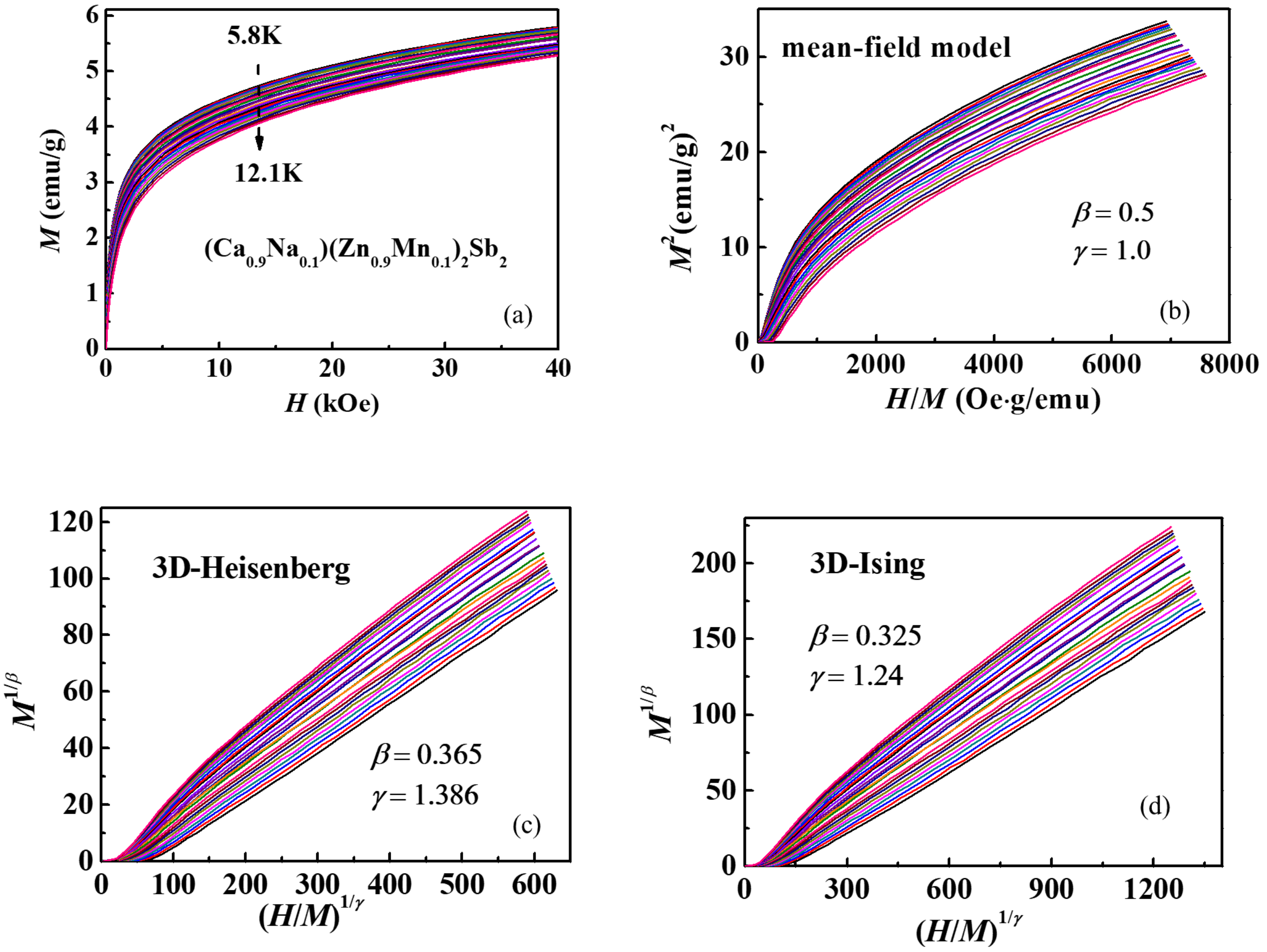

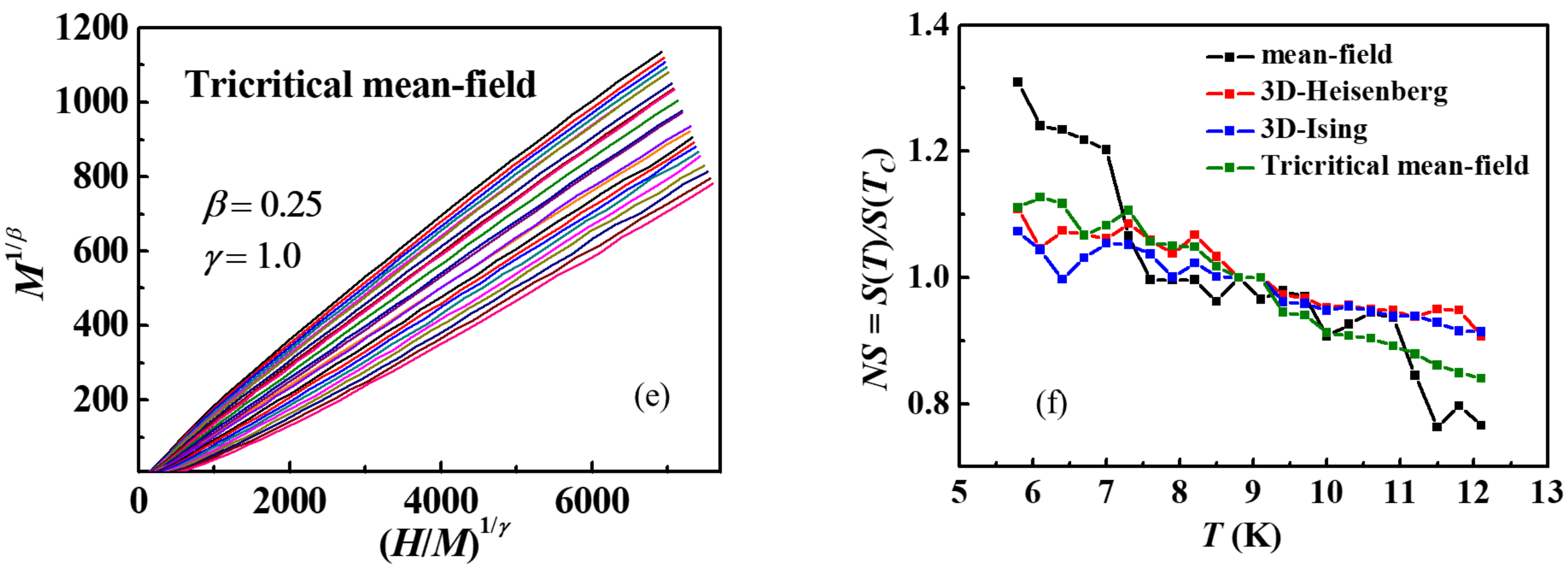

FIG. 4.

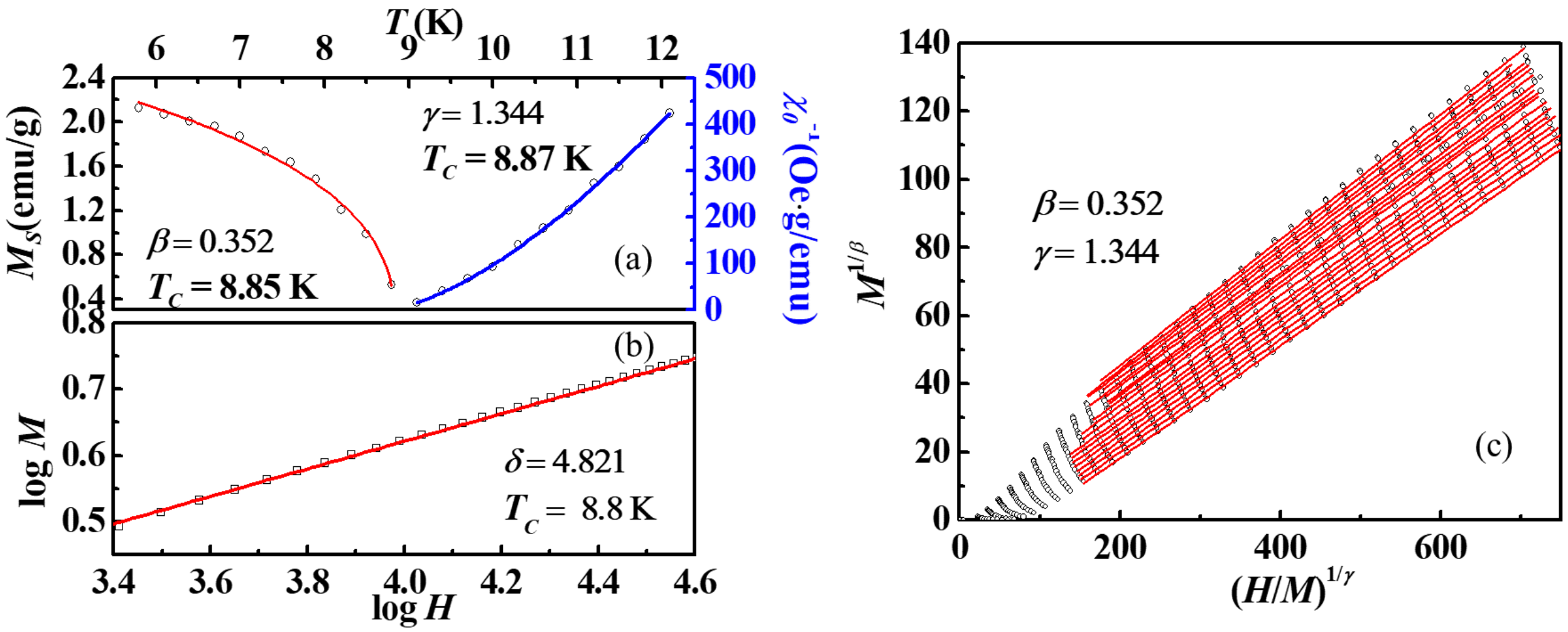

T(K)
6 7 8 9 10 11 12
2.4 2.0 1.6 1.2 0.8 0.4
Ms(emu/g)
γ = 1.344
TC = 8.87 K
β = 0.352
TC = 8.85 K
(a)
500 400 300 200 100 0
χ0-1(Oe·g/emu)
0.8 0.7 0.6 0.5
log M
(b)
δ = 4.821
TC = 8.8 K
3.4 3.6 3.8 4.0 4.2 4.4 4.6
log H
140 120 100 80 60 40 20 0
M1/β
β = 0.352
γ = 1.344
(c)
0 200 400 600
(H/M)1/γ

FIG. 5.

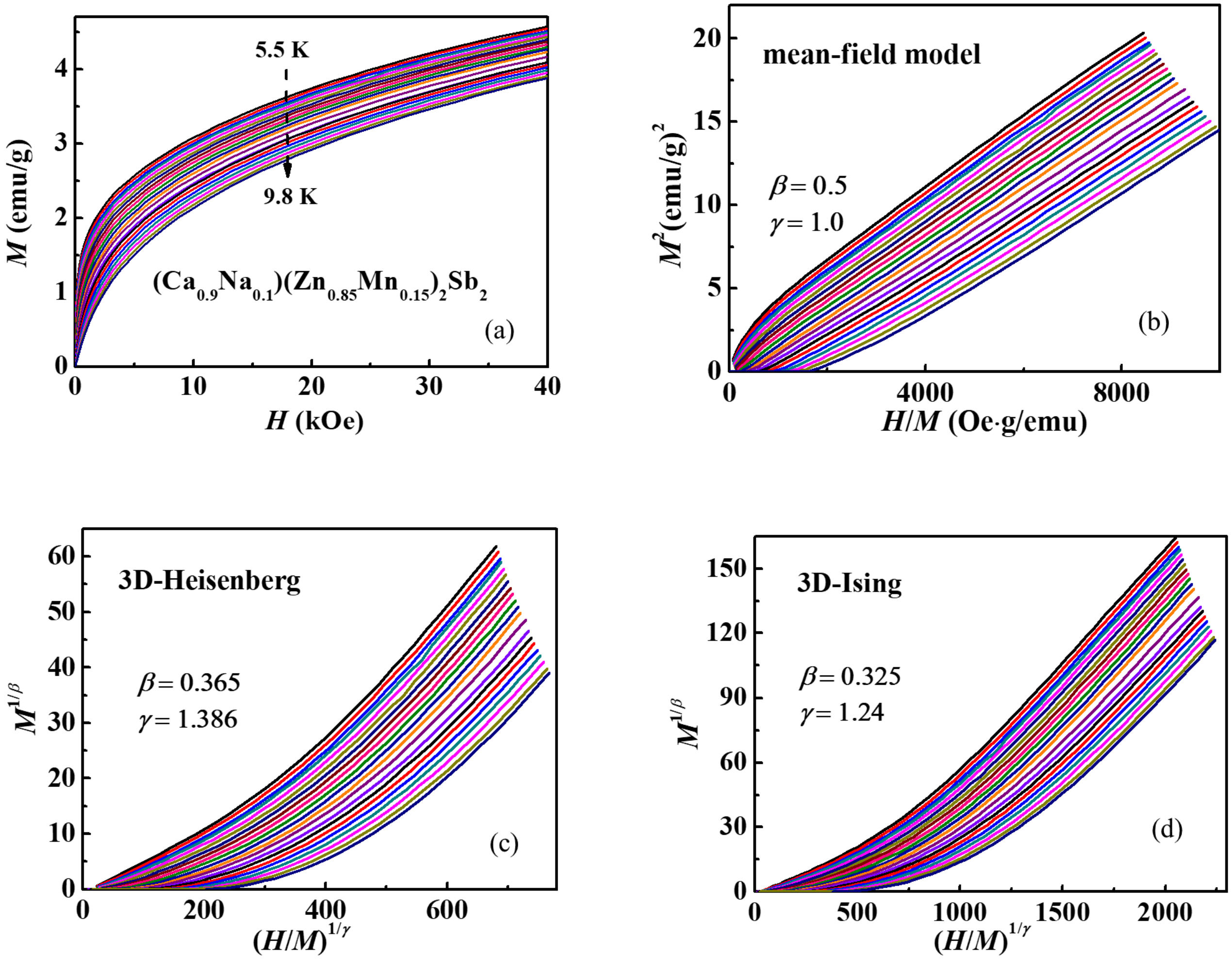

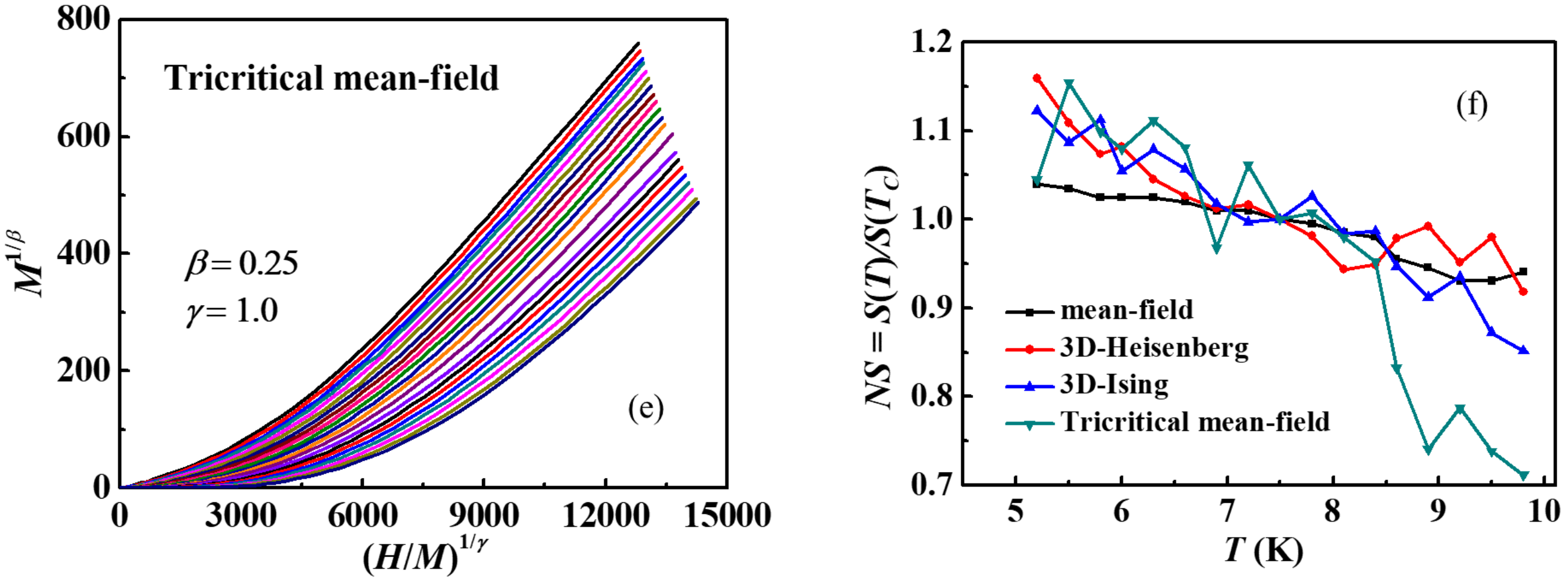

FIG. 6.

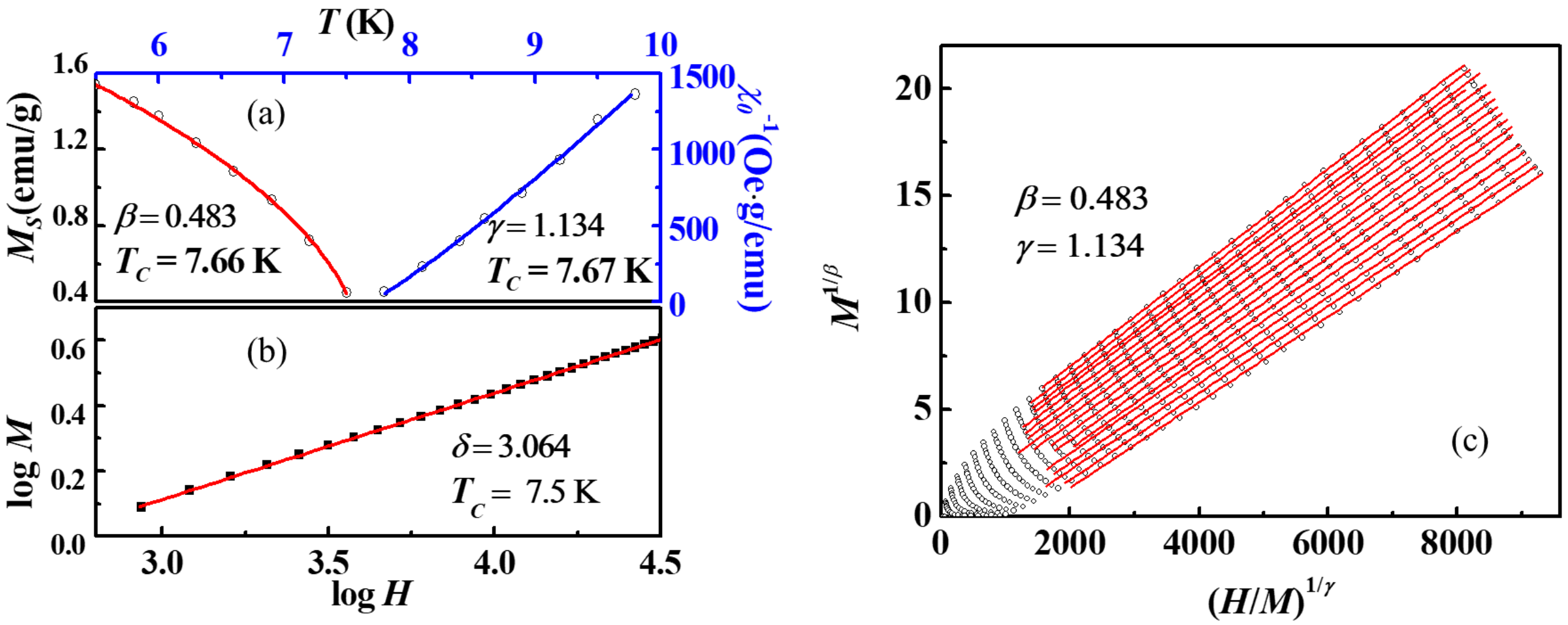

**Table I**

| *Materials* | *Technique* | $T_C$ (K) | $\beta$ | $\gamma$ | $\delta$ | *Ref.* |
|---|---|---|---|---|---|---|
| *Mean-field model* | Theory | | 0.5 | 1.0 | 3.0 | [37] |
| *3D-Heisenberg model* | Theory | | 0.365 ±0.003 | 1.386 ±0.004 | 4.80 ±0.04 | [37] |
| *3D-Ising model* | Theory | | 0.325 ±0.002 | 1.241 ±0.002 | 4.82 ±0.02 | [37] |
| *Tricritical mean-field model* | Theory | | 0.25 | 1 | 5 | [37] |
| $(Ca_{0.9}Na_{0.1})(Zn_{0.95}Mn_{0.05})_2Sb_2$ | MAP<br>**dM/dT*<br>(Widom) | 6.0<br>*6.4 | 0.322 ±0.010 | 1.385 ±0.011 | 5.078 ±0.021<br><br>(5.301) | This work |
| $(Ca_{0.9}Na_{0.1})(Zn_{0.90}Mn_{0.10})_2Sb_2$ | MAP<br>**dM/dT*<br>(Widom) | 8.8<br>*10.8 | 0.352 ±0.023 | 1.344 ±0.013 | 4.821 ±0.001<br><br>(4.818) | This work |
| $(Ca_{0.9}Na_{0.1})(Zn_{0.85}Mn_{0.15})_2Sb_2$ | MAP<br>**dM/dT*<br>(Widom) | 7.5<br>*7.9 | 0.483 ±0.012 | 1.134 ±0.043 | 3.064 ±0.006<br><br>(3.348) | This work |